\documentclass[a4paper]{aa}    
\usepackage[longnamesfirst,authoryear]{natbib}
\bibpunct{(}{)}{;}{a}{}{,} 
\usepackage{amsmath}
\usepackage{amssymb}
\usepackage{float}
\usepackage{epsfig}

\usepackage{txfonts}

\newcommand{\Mach}{\ensuremath{\mathcal{M}}}
\newcommand{\MSO}{MSO}
\newcommand{\ms}{{\ensuremath{\rm ms}}}
\newcommand{\up}[1][{}]{\ensuremath{u^{#1}_{\rm \hat p}}}
\newcommand{\cs}[1][{}]{\ensuremath{c^{#1}_{\rm s}}}
\newcommand{\ie}{{i.e.}}
\newcommand{\eg}{{e.g.}}
\newcommand{\etal}{{et al.}}
\newcommand{\cf}{{cf.}}
\newcommand{\rms}[1][{}]{\ensuremath{r^{#1}_{\rm ms}}}
\newcommand{\Thms}[1][{}]{\ensuremath{\Theta^{#1}_{\rm ms}}}

\begin{document}
   \title{Two-dimensional structure of thin transonic discs: theory and observational manifestations}
   \titlerunning{Two-dimensional structure of thin discs}
   \authorrunning{V. Beskin and A. Tchekhovskoy}

\author{V. Beskin
    \inst{1}
    \and
    A. Tchekhovskoy\inst{2}
    }

   \institute{Lebedev Physical Institute,\\
Leninskii prosp., 53, Moscow, 119991, Russia\\
              \email{beskin@lpi.ru}
         \and
             Moscow Institute of Physics and Technology,\\
             Institutskii per., 9, Dolgoprudny, 141700, Russia\\
             \email{chekhovs@lpi.ru}
             }

   \date{Received 5 July 2004 / Accepted 23 November 2004}

    \abstract{
We study the two-dimensional structure of thin transonic accretion
discs in the vicinity of black holes. We use the hydrodynamical
version of the Grad-Shafranov equation and focus on the region
inside the marginally stable orbit (\MSO), $r < \rms$. We show
that all components of the dynamical force in the disc become
significant near the sonic surface and especially important in the
supersonic region. Under certain conditions, the disc structure is
shown to be far from radial, and we review the affected disc
properties, in particular the role of the critical condition at
the sonic surface: it determines neither the accretion rate nor
the angular momentum in the accretion disc. Finally, we present a
simple model explaining quasi-periodical oscillations that have
been observed in the infrared and X-ray radiation of the Galactic
Centre.
   }

\maketitle

\section{Introduction}
\label{Sec.intro} The investigation of accretion flows near black
holes (BHs) is undoubtedly of great astrophysical interest.
Substantial energy is released near BHs, and general relativity
effects, attributable to strong gravitational fields, show up
there. Depending on external conditions, both quasi-spherical and
disc accretion flows can be realized.

The studies of thin disc accretion span more than three decades.
Many results were included in textbooks~\citep{sha83,lip92}.
\citet{lyn69} was the first to point out that supermassive BHs
surrounded by accretion discs could exist in galactic nuclei.
Subsequently, a theory for such discs was developed that is now
called standard model, or the model of the
$\alpha-$disc~\citep{sha72,sha73,nov73}.

According to standard model, if the accreting gas temperature is
low enough for the speed of sound $\cs$ to be much lower than the
Keplerian rotation velocity $v_{\rm K},$
matter forms a thin equilibrium disc and moves in nearly circular
orbits with the Keplerian velocity $v_{\rm K}(r)= (G M/r)^{1/2},$
where $M$ is the black hole mass and $G$ is the gravitational
constant. The disc thickness within this model is determined by
the balance of gravitational and accreting matter pressure forces,
\begin{equation}
H \approx r\frac{\cs}{v_{\rm K}}. \label{2}
\end{equation}
Minor friction between rotating gas layers leads to the loss of
angular momentum, so that the accreting matter gradually
approaches the compact object; this motion only slightly disturbs
the vertical balance\footnote{Let the \emph{vertical}, or
\emph{transversal}, direction be a $\theta$-direction in the
spherical system of coordinates with the half-line $\theta=0$
being aligned with the angular momentum vector of the infalling
matter. Then, the \emph{longitudinal} direction is a radial one.}
and is commonly considered a small perturbation. To account for
this phenomenon, standard model introduces a phenomenological
proportionality coefficient $\alpha_{\rm SS} \le 1$ between the
viscous stress tensor $t^r_{\varphi}$ and the pressure $P$,
$t^r_{\varphi} = \alpha_{\rm SS} P.$ As a result, the radial
velocity $v_r$ can be evaluated as
\begin{equation}
\frac{v_r}{v_{\rm K}} \approx \alpha_{\rm SS}\frac{\cs[2]}{v_{\rm
K}^2}, \label{v_r}
\end{equation}
so that for $\cs \ll v_{\rm K}$ the radial velocity is much lower
than both the Keplerian velocity and the speed of sound.

The foregoing is valid far from the compact object where the
relativistic effects are unimportant. As for the inner regions of
the accretion disc, the general relativity effects lead to at
least two new qualitative phenomena. First, there are no stable
circular orbits at small radii, $r<\rms$ for a non-rotating BH;
$\rms = 3r_{\rm g}$ is the radius of the marginally stable orbit
(\MSO) and $r_{\rm g} =2 GM/c^2$ is the BH gravitational radius.
This means that the accreting matter, that falls into the region
$r<\rms,$ rapidly approaches the BH horizon,\footnote{In the
dynamic time \hbox{$\tau_{\rm d} \sim [v_r(\rms)/c]^{-1/3}r_{\rm
g}/c$}.} no matter if the viscosity is present. Second, thin disc
accretion on to BHs is always transonic. This results from the
fact that, according to (\ref{v_r}), the flow is subsonic outside
the \MSO, $r>\rms$, while at the horizon $r = r_{\rm g}$ the flow
is to be supersonic \citep{bes97}. Therefore, the study of inner
accretion disc regions requires consistent analysis of transonic
flows.

Transonic accretion on to BHs has been the subject of many
research publications. Of crucial importance was the paper by
\citet{pac81} who formulated equations of motion in terms of
radial derivatives only: they used quantities averaged over the
disc thickness. Subsequently, many authors considered only such
one-dimensional models. In most cases, the model potential of
\citet{pac80}, \hbox{$\varphi_{\rm g} =-GM/(r-r_{\rm g}),$} which
simulates the properties of the strong gravitational field of a
centrally symmetric BH, was used
\citep{abr88,pap94,che97,nar97,art01}. The Schwarzschild and Kerr
metrics, that describe the gravitational field of axisymmetric
BHs,
were far more rare 
\citep{rif95,pei97,bel98,gam98a,gam98b}. As for the
two-dimensional structure, it was investigated mostly only
nume\-rical\-ly and on\-ly for thick discs
\citep{pap94,igu97,bal98,kro02}.

Therefore, for thin accretion discs basic results were obtained in
terms of the one-dimensional approach. However, the procedure for
averaging over the disc thickness, which is likely to be valid in
the region of stable orbits, in fact requires a more serious
analysis. In our view, it is the assumption that the transverse
velocity $v_\theta$ may always be neglected in thin accretion
discs, i.e. that the disc thickness is always determined by
(\ref{2}), that is the most debatable. This assumption is widely
used, explicitly or implicitly, virtually in all papers devoted to
thin accretion discs (see the papers cited above and
\citealp{abr81,cha96}).

\citet{abr97} pointed out the importance of allowing for the
transverse velocity in the inner accretion disc regions. They
argued that the transverse component of the dynamical force
has the same asymptotic near the equatorial plane as the
transversal components of the gravitational and pressure forces
(under the natural assumption $v_\theta \propto \cos\theta$ in the
limit $\cos\theta\rightarrow0$). Therefore, the fact that the
transversal velocity vanishes on the equator by no means implies
that $\left[(\mathbf{v \nabla})\mathbf{v}\right]_\theta$ may be
ignored. However, in the end the authors concluded that for thin
discs the dynamical term may still be disregarded so that the
accretion disc thickness can be determined from the balance of the
gravitational force and the pressure gradient up to the BH
horizon. We show that this conclusion is inapplicable for the
class of problems we consider below.

To refine the issue, it may be fruitful to recall the simplest
case of transonic accretion, namely ideal spherically symmetric
accretion \citep{bon52}\label{Pageref.Bondi}. Far outside the
sonic surface in such a flow the dynamical term is negligible in
comparison with the pressure gradient and the gravitational force.
However, all the terms become of the same order of magnitude near
the sonic surface. Moreover, after passing the sonic surface, the
motion of matter differs little from free-fall, therefore the
pressure gradient no longer affects the dynamics of accreting
matter. This makes it seem unreasonable to determine the disc
thickness from~(\ref{2}), at least in the supersonic region.

However, equation (\ref{2}) can be easily modified to account for
dynamical forces. Indeed, we have at the border of the disc
\begin{equation}
v_\Theta \approx \frac{dH}{dr} v_r, \label{Eq.utheta}
\end{equation}
where $\Theta = \pi/2 - \theta$. Now assuming linear dependence of
$v_\Theta$ on~$\Theta$ for $\Theta \ll 1,$ we obtain
\begin{equation}
 v_\Theta \approx
   \left(\frac{r}{H}\frac{dH}{dr}\right) v_r \Theta.
\end{equation}
With this in hand the dynamical term can be written as
\begin{gather}
 \left[(\mathbf{v \nabla})\mathbf{v}\right]_\Theta =
 v_r \frac{\partial (r v_\Theta)}{r\partial r} +
 v_\Theta \frac{\partial v_\Theta}{r\partial \Theta}
 \nonumber\\
 \approx \frac{v_r}{r}\frac{\partial}{\partial r}
 \left(r \frac{r}{H}\frac{dH}{dr} v_r\Theta\right)
 +  \left( \frac{r}{H}\frac{dH}{dr} v_r\Theta\right)
 \frac{\partial}{r\partial\Theta}
 \left( \frac{r}{H}\frac{dH}{dr} v_r\Theta\right).
 \label{Eq.dynam}
\end{gather}
Now, assuming $\Theta = H/r$ and using (\ref{Eq.dynam}), the
$\Theta$-component of the Euler equation, divided by $\left(-\rho
H/r^2\right),$ can be written as
\begin{equation}
 r v_r \frac{d}{dr}
    \left[
        \left(\frac{r}{H}\frac{dH}{dr}\right) v_r
    \right] +
 \left(\frac{r}{H} \frac{dH}{dr}\right)^2 v_r^2 =
 \cs[2]\frac{r^2}{H^2} - v_{\varphi}^2,
 \label{fullbalance}
\end{equation}
where the two terms in the l.h.s.\hbox{} correspond to the
dynamical force components $v_r \partial (r v_\theta)/(r\partial
r)$ and $v_\theta \partial v_\theta/(r\partial \theta)$
respectively. The r.h.s.\hbox{} of (\ref{fullbalance}) corresponds
to the discrepancy between the centrifugal force
$v_{\varphi}^2\tan\Theta\,/r$ and the pressure force $\partial
P/(r\partial\Theta)$ whose absolute value is evaluated as $P/H
\sim \rho\cs[2]/H.$ This discrepancy was postulated to vanish in
(\ref{2}).

Accounting for the dynamical forces, equation~(\ref{fullbalance})
is a simple qualitative analogue of second-order Grad-Shafranov
equation~(\ref{GSfull}) and allows significant disc thickness
deviations from the value given by conventional
prescription~(\ref{2}). For instance, nondimensionalizing
equation~(\ref{fullbalance}) yields $\delta r,$ the characteristic
radial scale of change of $H$ near the sonic surface $r=r_*$,
$\delta r(r_*) \sim H(r_*).$ Consistently, by (\ref{Eq.utheta}) it
follows that $v_\theta \sim v_r$ which indeed indicates a rapid
change of the disc thickness. Naturally, in the region of stable
orbits in a similar way we get $\delta r(r>\rms) \sim r$ and
$v_\theta\sim v_r H/r \ll v_r$ which justifies the usage of
conventional prescription (\ref{2}) outside the \MSO.

The existence of the small longitudinal scale \hbox{$\delta r(r_*)
\ll r_{\rm g}$} is one of the key results of this paper. Note also
that since (\ref{fullbalance}) is a second-order differential
equation for $H,$ there in fact exist two additional degrees of
freedom, frequently overlooked before, and only one of the two is
fixed by the critical condition at the sonic surface. This hints
that the critical condition determines neither the accretion rate
nor the angular momentum in the disc but only imposes a
restriction on the function $H(r),$ \ie\ on the form of the flow,
near the sonic surface.

In this paper we purposefully consider extreme boundary conditions
with radial inflow velocity at the \MSO\ being far smaller than
the speed of sound. This allows us to perform the study of the
problem analytically. Moreover, the results turn out to be almost
independent of the radial inflow velocity, and in the end the
observational predictions of our model depend only on the the
black hole mass and its angular momentum and are totally
independent of the inflow velocity and the speed of sound.

Although our model might not be realized in Nature as is, it~-- as
any other analytical model~-- bolsters our fundamental
understanding of the physical process. Namely, this model offers
an insight into the fundamental properties we might lose if we do
not allow for the transversal velocity in thin accretion discs.
This is the first step to understanding the physics of the
critical condition at the sonic surface and its influence on the
global disc structure.

In the next section we introduce the reader to the hydrodynamical
version of the Grad-Shafranov equation in the Schwarzschild metric
as well as derive some useful relations. In sections
\ref{Sec.SubsonicFlow} and \ref{Sec.TransFlow} we sequentially
study the subsonic and the sonic surface regions. Section
\ref{Sec.SupersonicFlow} presents the study of the supersonic flow
for the case of a spinning BH, \ie\ Kerr metric, followed by the
discussion of its applications to the explanation of
quasi-periodical oscillations detected in the infrared and X-ray
observations of the Galactic Centre (section
\ref{Section.Observ}).

\section{Basic equations}
\label{Sec.BasicEqs} We consider thin disc accretion on to a BH in
the region where there are no stable circular orbits. As argued
above, the contribution of viscosity should no longer be
significant here. Hence we may assume that ideal hydrodynamics
approach is suitable well enough for describing the flow structure
in this inner area of the accretion disc.
Below, unless specifically stated, we consider the case of a
non-spinning BH, \ie\ use the Schwarzschild metric, and use a
system of units with $c=G=1.$ We measure radial distances in the
units of $M$, the BH mass.

In \hbox{Boyer-Lindquist} coordinates the Schwar\-z\-schild metric
is~\citep{lan87a}
\begin{equation}
 {\rm d}s^{2}=-\alpha^{2}{\rm d}t^{2}
 + g_{ik}{\rm d}x^{i}{\rm  d}x^{k}, \label{a1}
\end{equation}
where
\begin{gather}
 \alpha^2 = 1-2/r, \quad g_{rr} = \alpha^{-2}, \quad
 g_{\theta \theta} = r^2, g_{\varphi \varphi} = \varpi^{2} =
 r^2\sin^2\theta.\label{a2}
\end{gather}
Here and below indices without caps denote vector components with
respect to the coordinate basis in `absolute'\hbox{}
three-dimensional space, and indices with caps denote their
physical components. The $\nabla_k$ symbol always represents
covariant derivative in `absolute'\hbox{} three-dimensional space
with $g_{ik}$ metrics (\ref{a2}).

We reduce our discussion to the case of axisymmetric stationary
flows. This makes it possible to introduce a stream function
$\Phi(r,\theta)$ which defines the physical poloidal $4$-velocity
component $\mathbf{u}_{\rm \hat p} \equiv u_{\hat r}
\mathbf{e}_{\hat r} + u_{\hat\theta} \mathbf{e}_{\hat \theta}$ as
\begin{equation}
\alpha n\mathbf{u}_{\rm \hat p}=
\frac{1}{2\pi\varpi}(\mathbf{\nabla}\Phi\times\mathbf{e}
_{\hat\varphi}), \label{defPhi}
\end{equation}
where $n$ is the particle concentration in the comoving reference
frame. 
The surfaces $\Phi(r,\theta)=$ const determine the streamlines of
the matter ($\Phi = 0$ corresponds to the streamline $\theta =
0$).

For an ideal flow there are three integrals of motion conserved
along the streamlines, namely entropy,
\begin{gather}
 S = S(\Phi), \label{int_s}\\
 \intertext{energy flux,}
 E(\Phi) = \mu\alpha\gamma, \label{int_e}\\
 \intertext{and the $z$-component of angular momentum,}
 L(\Phi) = \mu\varpi u_{\hat\varphi}, \label{int_l}
\end{gather}
where $\mu=(\rho_{m}+P)/n$ is relativistic enthalpy; $\rho_{m}$ is
internal energy density, $P = n T$ is pressure.

As a result the relativistic Euler equation \citep{fro98}\hfill
\begin{gather}
nu^{b}\nabla_{b}(\mu u_{a})+\nabla_{a}P - \mu n(u_{\hat
\varphi})^{2} \frac{1}{\varpi}\nabla_{a}\varpi+\mu
n\gamma^{2}\frac{1}{\alpha}\nabla_{a}\alpha = 0, \label{euler}
\end{gather}
in which indices $a$ and $b$ take on $r$ and $\theta$ values, can
be rewritten in the form of the Grad-Shafranov scalar equation for
$\Phi(r,\theta)$ \citep{bes97},
\begin{align}
 &-\Mach^{2}\Bigl[\frac{1}{\alpha}\nabla_{k}
 \left(\frac{1}{\alpha\varpi
 ^{2}}\nabla^{k}\Phi\right)\notag\\
 &+\frac{1}{\alpha^{2}\varpi^{2}(\mathbf{\nabla}
 \Phi)^2}\frac{\nabla^{i}\Phi\nabla^{k}\Phi\nabla_{i}\nabla_{k}\Phi}{D}
 \Bigr]
 +\frac{\Mach^{2}\nabla'_{k}F\nabla^{k}\Phi}{2\alpha^{2}
 \varpi^{2}(\mathbf{\nabla}\Phi)^{2}D}\notag\\
 &+\frac{64\pi^{4}}{\alpha^{2}\varpi^{2}\Mach^{2}}
 \left(\varpi^{2}E\frac{{\rm d}E}{{\rm d}\Phi}
 -\alpha^{2}L\frac{{\rm d}L}{{\rm d}\Phi}\right)
 -16\pi^{3}nT\frac{{\rm d}S}{{\rm d}\Phi}=0, \label{GSfull}
\end{align}
where the $\nabla'_{k}$ derivative acts on all the variables
except the quantity $\Mach$. The equation above contains three
integrals $E(\Phi)$, $L(\Phi)$, and $S(\Phi).$ Here
\begin{align}
&D=-1+\frac{1}{u^{2}_{\rm p}}\frac{c^{2}_{s}}{1-c^{2}_{s}},
\label{b9} \\
&F=\frac{64\pi^{4}}{\Mach^{4}}\left[\varpi^{2}E^{2}
-\alpha^{2}L^{2}-\varpi^{2}\alpha^{2}\mu^{2}\right], \label{b10}\\
\intertext{and the thermodynamical function $\Mach^{2%
}$ is defined as} &\Mach^{2}=\frac{4\pi\mu}{n}. \label{b7}
\end{align}
Equation (\ref{GSfull}) contains the only singular surface -- the
sonic surface which is defined by the condition $D=0$. At this
surface the equation changes its type from elliptical to
hyperbolical.

To close the system, we need to supply the Grad-Shafranov equation
with the
relativistic Bernoulli equation, \hbox{$\up[2] =
\gamma^2-u_{\hat\varphi}^2 - 1$}, which now becomes
\begin{align}
\up[2] &= \frac{E^2-\alpha^2L^2/\varpi^2
-\alpha^2\mu^2}{\alpha^2\mu^2}. \label{up2full}
\intertext{After using definition (\ref{defPhi}) we reach} %
E^{2} &= \alpha^{2}\mu^{2}+\frac{\alpha^2}{\varpi^2}L^2+
\frac{\Mach^4}{64\pi^{4}\varpi^2}(\mathbf{\nabla}\Phi)^2.
\label{b6}
\end{align}

For the sake of simplification we adopt the polytropic equation of
state $P = k(S)n^{\Gamma}$ so that temperature and sound velocity
can be written as~\citep{sha83}
\begin{equation}
T = k(S)n^{\Gamma - 1}; \qquad \cs[2] =
\frac{\Gamma}{\mu}k(S)n^{\Gamma - 1}. \label{Tc}
\end{equation}
Since the disc is thin, $\cs \ll 1$ (see (\ref{2})). Therefore we
can write $\mu = m_{\rm p} + m_{\rm p}W,$ where $W =
\cs[2]/(\Gamma - 1)$ is non-relativistic enthalpy and $m_{\rm p}$
is particle mass. For an ideal gas with \hbox{$\Gamma = {\rm
const}$} the function $k(S)$ can be shown to have a quite definite
form
\begin{equation} k(S) = k_0\exp{[(\Gamma -
1)S]}, \label{ks}
\end{equation}
which for the case of the polytropic equation of state can be
obtained from (\ref{Tc}) and the thermodynamical relationship
${\rm d}P = n{\rm d}\mu - nT{\rm d}S$.

Note that equation (\ref{b6}) enables us to express the quantity
$\Mach$ and together with it all the other thermodynamical
quantities in terms of the potential $\Phi(r,\theta)$ and the
three motion invariants. Consequently, equation (\ref{GSfull})
does contain the only unknown function $\Phi(r,\theta).$

\section{Subsonic flow}
\label{Sec.SubsonicFlow}
In this section we show that the role of the dynamical terms becomes dominant with
approach to the sonic surface $r = r_*(\theta)$. The problem of
passing through the sonic surface and the supersonic flow
structure are dealt with later in this paper. Here we limit our
discussion to the subsonic region only, where the poloidal
velocity $\up$ is far lower than the sound velocity~$\cs$ (as we
assume in Sec.\hbox{}~\ref{Sec.intro}, the condition
$\up[2]/\cs[2] \ll 1$ holds at the \MSO, and in this section we
always neglect the terms of the order of $\up[2]/\cs[2]$ or
higher). We can simplify Grad-Shafranov hydrodynamic
equation~(\ref{GSfull}) in this limit
by neglecting terms proportional to $D^{-1}$:
\begin{gather}
-\Mach^{2}\frac{1}{\alpha}\nabla_{k}
\left(\frac{1}{\alpha\varpi^{2}}\nabla^{k}\Phi\right)\notag\\
+\frac{64\pi^{4}}{\alpha^{2}\varpi^{2}\Mach^{2}}
\left(\varpi^{2}E\frac{{\rm d}E}{{\rm d}\Phi}
-\alpha^{2}L\frac{{\rm d}L}{{\rm d}\Phi}\right)
-16\pi^{3}nT\frac{{\rm d}S}{{\rm d}\Phi}=0. \label{main'}
\end{gather}
The resulting equation describes the subsonic flow and is
elliptical. Hence, it contains no critical surfaces, and our
problem requires five boundary conditions: three conditions
determine the integrals of motion and the two remaining ones are
the boundary
conditions for the second-order Grad-Shafranov equation.

Following Sec.\hbox{}~\ref{Sec.intro}, we assume that the
$\alpha$-disc theory holds outside the \MSO. We adopt the flow
velocity components, which this theory yields on the \MSO\
$r=\rms,$\footnote{A nearly parallel inflow with small radial
velocity $v_r$ (\ref{v_r}).} as the first three boundary
conditions for our problem. For the sake of simplicity we consider
the radial velocity, which is responsible for the inflow, to be
constant at the surface $r = \rms$ and equal to $u_0$ and the
toroidal velocity to be exactly equal to that of a free
particle revolving at $r=\rms$:%
\footnote
 {
 For a free particle revolving at $r=\rms$ around a non-spinning
 BH we have \hbox{$u_{\hat\varphi}(\rms) = 1/\sqrt{3}$},
 \hbox{$\alpha_0 = \alpha(\rms) = \sqrt{2/3}$}, \hbox{$\gamma_0 = \gamma(\rms) =
 \sqrt{4/3}$} \citep{lan87a}.\label{Ftn.MS}
 }
\begin{align}
u_{\hat r}(\rms,\Theta)      &=  -u_0, \label{ur} \\
u_{\hat \Theta}(\rms,\Theta) &= \Theta u_0, \label{utet}\\
u_{\hat\varphi}(\rms,\Theta) &= 1/\sqrt{3}, \label{uphi}
\end{align}
where $\left|u_{\hat \Theta}\left(\rms,\Theta\right)\right|  \ll
\left|u_{\hat r}\left(\rms,\Theta\right)\right|$ accounts for the
fact that the flow is parallel\footnote{Note that the parallel
flow is not exactly the same as the radial one.} at the \MSO. The
angular variable $\Theta = \pi/2 -\theta$ is counted off from the
equator in the vertical direction.

For the sake of convenience, we also introduce another angular
variable $\Thms = \Thms(\Phi(r, \Theta))$, {a Lagrange coordinate
of streamlines.} This is a function of streamlines. For each
streamline, it gives the streamline's \hbox{$\Theta$-coordinate}
at the \MSO\ $r=\rms.$ In other words, both points $(r, \Theta)$
and $(\rms, \Thms)$ belong to the same streamline. Further, noting
that $\Thms(\rms, \Theta) \equiv \Theta$ and using (\ref{defPhi})
and (\ref{ur}), we arrive at
\begin{equation}
{\rm d}\Phi = 2\pi \alpha_0 \rms[2] n(\rms, \Thms)u_0\cos\Thms
{\rm d}\Thms. \label{d3}
\end{equation}

Next, we assume that the sound velocity is constant at the surface
$r = \rms;$ this yields the fourth boundary condition
\begin{equation}
\cs(\rms,\Theta) = c_0. \label{c0}
\end{equation}
In the case of polytropic equation of state this means that both
the temperature and the relativistic enthalpy are also constant at
the surface $r=\rms$, \ie:
\begin{align}
T(\rms, \Theta) &= T_0,\\
\mu(\rms, \Theta) &= \mu_0.
\end{align}
Therefore boundary conditions (\ref{ur}) \nolinebreak --
\nolinebreak (\ref{uphi}) and (\ref{c0}) directly determine the
invariants $E(\Thms)$ and $L(\Thms)$,
\begin{align}
E(\Thms) & = \mu_0 e_0,
\label{d1} \\
L(\Thms) & = \mu_0 l_0 \cos\Thms, \label{d2}
\end{align}
where $e_0 = \alpha_0\gamma_0 = \sqrt{8/9}$ and $l_0 =
u_{\hat\varphi}(\rms)\rms = \sqrt{3}r_{\rm g} = 2 \sqrt{3}$ (see
footnote \ref{Ftn.MS}). 

Condition $E={\rm const}$ (\ref{d1}) allows us to rewrite the
Grad-Shafranov equation in an even simpler form,
\begin{gather}
\frac{\partial^2 \Phi}{\partial r^2}
+\frac{\cos\Theta}{\alpha^2r^2}\frac{\partial}{\partial\Theta}
\left(\frac{1}{\cos\Theta}\frac{\partial\Phi}{\partial
\Theta}\right)  \notag\\
= -4\pi^2n^2\frac{L}{\mu^2}\,\frac{{\rm d}L}{{\rm d}\Phi} -
4\pi^{2}n^2r^2\cos^2\Theta\frac{T}{\mu}\,\frac{{\rm d}S}{{\rm
d}\Phi}. \label{main}
\end{gather}
At $r=\rms$ the r.h.s.\hbox{} of equation (\ref{main}) can be
shown to describe the transversal balance of the pressure force
and the effective potential, whereas the l.h.s.\hbox{} corresponds
to the dynamical term $(\mathbf{v\nabla})\mathbf{v}$ which is
$u_0^2/c_0^2$ times larger than each of the terms in the
r.h.s.\hbox{} and thus may be dropped \citep{bes02}. It is natural
therefore to choose entropy $S(\Phi)$ from the condition of the
transversal balance of forces at the surface $r = \rms,$
\begin{equation}
\rms[2]\cos^2\Thms\frac{{\rm d}S}{{\rm d}\Thms} =
-\frac{\Gamma}{c_0^2}\,\frac{L}{\mu_0^2}\,\frac{{\rm d}L}{{\rm
d}\Thms}, \label{sr0}
\end{equation}
where the value of $L(\Thms)$ is determined by (\ref{d2}). This
yields the last, fifth, boundary
condition
\begin{equation}
S(\Thms) = S(0) - \frac{\Gamma}{3c_ 0^2}\ln(\cos\Thms), \label{sp}
\end{equation}
whence owing to (\ref{ks}) 
we have at $r = \rms$ the standard shape of concentration,
\begin{equation}
n(\rms, \Theta) \approx n_0
\exp\left(-\frac{\Gamma}{6c_0^2}\Theta^2\right) \label{np}
\end{equation}
(to be exact, $n(\rms,\Theta) = n_0(\cos\Theta)^{\Gamma/3c_0^2}$).
Now with the help of~(\ref{d3}), invariants (\ref{d1}),
(\ref{d2}), and (\ref{sp}) can be expressed in terms of $\Phi$,
therefore the Grad-Shafranov equation does contain the only
unknown function $\Phi(r,\Theta)$.

\begin{figure}
    \begin{center}
    \epsfig{figure=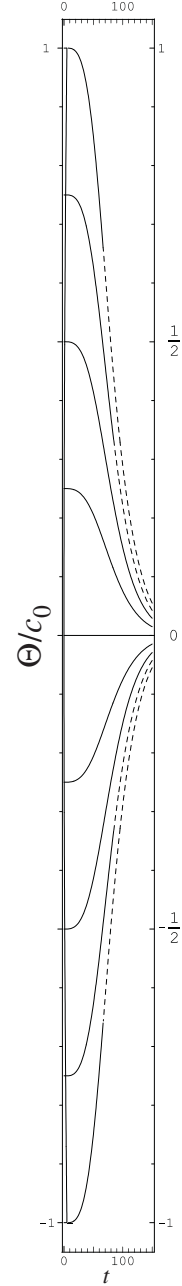,height=0.7 \textheight}
    \caption{The disc streamlines profile (real scale)
obtained from the numerical computation of equation (\ref{main})
for $c_0 = 10^{-2}$ and $u_0 = 10^{-5}$. Solid lines correspond to
the range of variables $\up[2]/\cs[2] < 0.2,$ where the solution
to approximate equation~(\ref{main}) should not differ greatly
from the solution to complete equation~(\ref{GSfull}).
Extrapolation of the solution to the sonic surface region is shown
with dashed lines. In the figure we use the radial variable
$t=(\rms-r)/(\rms u_0)\colon$ for the above choice of $u_0$ and
$c_0,$ the point $t=200$ corresponds to $r = 2.994 r_{\rm g}.$}
     \label{Fig.Subsonic}
     \end{center}
\end{figure}

\begin{figure}
  \begin{center}
    \epsfig{figure=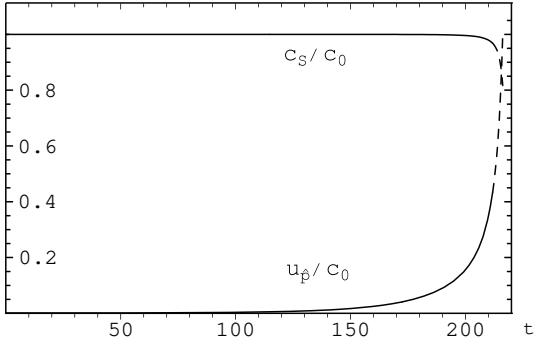,width=0.8 \linewidth}
         \caption{Profiles of the poloidal velocity $\up$ and sound
velocity $\cs$ at the equator for $c_0 = 10^{-2}$ and $u_0 =
10^{-5}$. Solid lines correspond to the same range of parameters
as in Fig.\hbox{}~\ref{Fig.Subsonic}
}
     \label{Fig.VelocityProfiles}
   \end{center}
\end{figure}

The disc streamlines profile obtained from numerical computation
of equation~(\ref{main}) is shown in
Fig.\hbox{}~\ref{Fig.Subsonic}; the profiles of radial velocity
and sound speed are shown in
Fig.\hbox{}~\ref{Fig.VelocityProfiles}. In the subsonic region,
$r_* < r \le \rms \equiv 3 r_{\rm g}$, the disc thickness rapidly
diminishes, and at the sonic surface we have $H(r_*) \approx
u_0/c_0 H(\rms)$ so that we cannot neglect the dynamical force
there~\citep{bes02}.

This can be readily derived from qualitative consideration as
well. For $\cs \ll 1$, \ie\ for non-relativistic temperature, we
obtain from (\ref{up2full}) and (\ref{Tc}),
\begin{equation}
\up[2]  = u_0^2 +w^2 + \frac{2}{\Gamma - 1}\left(c_0^2 -
\cs[2]\right) + \frac{1}{3}\left(\Thms[2]-\Theta^2\right) + \dots
\label{uup}
\end{equation}
The quantity
\begin{equation}
w^2(r) = \frac{e_0^2-\alpha^2l_0^2/r^2-\alpha^2}{\alpha^2} \equiv
\frac{1}{\alpha^2}\frac{(6-r)^3}{9r^3} \label{w2}
\end{equation}
is the poloidal four-velocity of a free particle having zero
poloidal velocity at the \MSO.
Assuming $\up = \cs = c_*$ and neglecting $w^2$ in (\ref{w2}), we
find the velocity of sound $c_*$ in the sonic point $r = r_*$,
$\Theta = 0$:
\begin{equation}
c_* \approx \sqrt{\frac{2}{\Gamma+1}}\,c_0. \label{cc}
\end{equation}
Since entropy $S$ remains constant along the streamlines, the gas
concentration $n_*$ at the sonic surface slightly
differs\footnote{Note that the concentration, certainly, changes
from one streamline to another.} from the gas concentration at the
\MSO. In other words, the subsonic flow can be considered
incompressible to a zeroth approximation.\footnote{The same is
true for spherically symmetric Bondi accretion \citep{sha83}.} It
is important that this conclusion holds not only in the equatorial
plane because the additional term $1/3(\Thms[2] - \Theta^2)$
in~(\ref{uup}) is also of the order of $c_0^2$ for the range of
angles corresponding to the representative disc thickness, $\Theta
\lesssim \Theta_{\rm disc} \sim c_0$ (see (\ref{np})).
%
Since the density remains almost constant and the poloidal
velocity increases from $u_0$ to $c_* \sim c_0$, \ie\ changes over
several orders of magnitude for $u_0^2 \ll c_0^2,$ the disc
thickness $H$ should change in the same proportion due to the
continuity equation:
\begin{equation}
H(r_*) \approx \frac{u_0}{c_0}H(\rms). \label{compr}
\end{equation}
Thus, purely radial motion approximation is inapplicable in the
vicinity of the sonic surface, and both components of the
dynamical force become comparable to the pressure gradient near
the sonic surface \citep{bes02},
\begin{equation}
\frac{u_{\hat \Theta}}{r}\, \frac{\partial u_{\hat
\Theta}}{\partial \Theta} \approx u_{\hat r}\frac{\partial u_{\hat
\Theta}}{\partial r} \approx \frac{\nabla_{\hat \Theta}P}{\mu}
\approx \frac{c_0^2}{u_0^2} \frac{\Theta}{r},
\end{equation}
with the position of the sonic surface $r_*$ being determined by
\begin{equation}
\rms - r_* = \Lambda u_0^{2/3}\rms, \label{r*}
\end{equation}
where the logarithmic factor
\begin{equation}
 \Lambda = \left(\frac{3}{2}\right)^{2/3} 
		\left[
		   \ln\left(\frac{c_0}{u_0}\right)
		\right]^{2/3} \approx 5 - 7,
\end{equation}
and the radial logarithmic
derivative of concentration near the sonic surface being estimated
as
\begin{equation}
\eta_1 = (r/n)\left(\partial n/\partial r\right) \sim u_0^{-1}.
\label{logder}
\end{equation}

In other words, if there appears a nonzero vertical velocity
component, the dynamical term $({\bf v}\nabla){\bf v}$ cannot be
neglected in the vertical force balance near the sonic surface.
This property is apparently valid for arbitrary radial velocities
of the flow, \ie\ even if the transverse contraction of the disc
is not so pronounced.

\section{Transonic flow}
\label{Sec.TransFlow}

In order to verify our conclusions we consider the flow structure
in the vicinity of the sonic surface in more detail. Since the
smooth transonic flow is analytical at a singular point $r = r_*,
\Theta = 0$~\citep{lan87b}, we can
write
\begin{align}
 n & =  n_*\left(1+\eta_1h+
     \frac{1}{2}\eta_3\Theta^2+\dots\right),
\label{expansionn}\\
 \Thms &=  a_0\left(\Theta+a_1h\Theta+\frac{1}{2}a_2h^2\Theta+
            \frac{1}{6}b_0\Theta^3 + \dots
            \right),
\label{expansionthetam}
\end{align}
where $h = (r-r_*)/r_*$. Here we assume that all the three
invariants $E,$ $L,$ and $S$ are given by boundary conditions
(\ref{d1}), (\ref{d2}), and (\ref{sp}) respectively. Hence, the
problem needs only one more boundary condition.

\begin{figure}
  \begin{center}
    \epsfig{figure=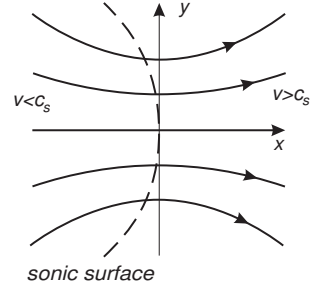,width=4cm}
    \caption{Schematics of thin disc stre\-am\-lin\-es pro\-file around the
    sonic point. The flow has the form of the standard nozzle.
    Here \hbox{$x = -h$}, and \hbox{$y = \Theta$}.}
     \label{Fig.transonic}
   \end{center}
\end{figure}

Full stream equation~(\ref{GSfull}) may be written as
\begin{align}
&-\frac{1}{\alpha}\nabla_{k}
\left(\frac{1}{\alpha\varpi^{2}}\Mach^{2}\nabla^{k}\Phi\right)\notag\\
&+\frac{64\pi^{4}}{\alpha^{2}\varpi^{2}\Mach^{2}}
\left(\varpi^{2}E\frac{{\rm d}E}{{\rm d}\Phi}
-\alpha^{2}L\frac{{\rm d}L}{{\rm d}\Phi}\right)
-16\pi^{3}nT\frac{{\rm d}S}{{\rm d}\Phi}=0. \label{GSfull'}
\end{align}
Neglecting spatial derivatives of $\mu$ here (this can be done in
the vicinity of the sonic surface for $c_0\ll 1$ as is shown
later) we get an equation with a form similar to (\ref{main}), but
this time the equation can be used in the vicinity of the sonic
surface. Now comparing the appropriate coefficients in Bernoulli
(\ref{up2full}) and full stream equation (\ref{GSfull'}) we obtain
neglecting terms of the order of $\varepsilon = u_0^2/c_0^2$ (see
the Appendix for the exact values),
\begin{align}
a_0 &= \left(\frac{2}{\Gamma+1}\right)^%
        {(\Gamma+1)/2(\Gamma-1)}\frac{c_0}{u_0},
\label{a_0}\\
a_1 &= 2 + \frac{1-\alpha_*^2}{2\alpha_*^2} \approx 2.25,
\label{a_1} \\
a_2 &= -(\Gamma+1)\eta_1^2,
\label{a_2}\\
b_0 &=
 \left(\frac{\Gamma + 1}{6}\right) \frac{a_0^2}{c_0^2}, \\
\eta_3 &= -\frac{2}{3} (\Gamma + 1)\eta_1^2 -
\left(\frac{\Gamma-1}{3}\right)\frac{a_0^2}{c_0^2},
  \label{eta_3}
\end{align}
where $\alpha_*^2 = \alpha^2(r_*) \approx 2/3$.
As we see, coefficients (\ref{a_0})~--~(\ref{eta_3}) are expressed
through radial logarithmic derivative $\eta_1$ (which we specify
according to (\ref{logder}) as the fourth boundary condition).
They have clear physical meaning. So, $a_0$ gives the compression
of streamlines: $a_0 = H(\rms)/H(r_*)$. In agreement with
(\ref{compr}) we have $a_0 \approx c_0/u_0$. Further, $a_1$
corresponds to the slope of the streamlines with respect to the
equatorial plane. As $a_1 > 0$, the compression of streamlines
finishes somewhere before the sonic surface, so inside the sonic
radius $r < r_*$ the streamlines diverge. On the other hand,
because $a_1 \ll u_0^{-1}$, for $r = r_{*}$ the divergency is
still very weak. Hence in the vicinity of the sonic surface the
flow has the form of the standard nozzle (see
Fig.~\ref{Fig.transonic}).
Finally, since $a_2 \sim \eta_3 \sim b_0
\sim u_0^{-2}$, 
the transverse scale of the transonic region $H(r_*)$ is the same
as the longitudinal one.
This means that the transonic region is essentially
two-dimensional (a well-known fact for a nozzle, see \eg\
\citet{lan87b}), and it is impossible to analyze it within the
standard one-dimensional approximation.

Since the transonic flow in the form of a nozzle has longitudinal
and transversal scales of one order of magnitude, 
 near the sonic surface we have $\delta r_\parallel
\approx \delta r_\bot,$ \ie\
\begin{equation}
\delta r_\parallel \approx H(r_*).
\end{equation}
Hence, for thin discs
this longitudinal scale is always much smaller than the distance
from the BH, $\delta r_\parallel/r_* \approx H(r_*)/r_* \ll 1$.
Only by taking the transversal velocity into account do we retain
the small longitudinal scale $\delta r_\parallel \ll r_{\rm g}.$
This scale is left out during the standard one-dimensional
approach.

We stress that taking the dynamical force into account is indeed
extremely important. This is because, unlike zero-order standard
disc thickness prescription (\ref{2}), the Grad-Shafranov equation
has second order derivatives, \ie\ contains two additional degrees
of freedom. This means that the critical condition only fixes one
of these degrees of freedom (\eg\ imposes some limitations on the
form of the flow) rather than determines the angular momentum of
the accreting matter.

Connecting the sonic characteristics $\eta_1 = \eta_1(r_*)$ with
the physical boundary conditions on the \MSO\ $r = \rms$ is rather
difficult (for this we have to know all the expansion coefficients
in~(\ref{expansionn}) and~(\ref{expansionthetam})). In particular,
we cannot formulate the restriction on five boundary conditions
(\ref{ur})~-- (\ref{uphi}), (\ref{c0}), and~(\ref{sp}) resulting
from the critical condition on the sonic surface. Nevertheless,
estimate (\ref{logder}) of $\eta_1$ makes sure that we know the
parameter $\eta_1$ to a high enough accuracy. Then, according to
(\ref{a_0})~-- (\ref{eta_3}), all other coefficients can be
determined exactly.

Using expansions (\ref{expansionn}) and (\ref{expansionthetam}),
one can obtain all other physical parameters of the transonic
flow. In particular, we have
\begin{align}
 \up[2] &= c_*^2
   \left[1-2\eta_1 h+
         \frac{1}{6}(\Gamma - 1) \, \frac{a_0^2}{c_0^2}\Theta^2
                  + \frac{2}{3}(\Gamma + 1)\eta_1^2\Theta^2
   \right], \nonumber \\
 \cs[2] &= c_*^2
   \biggl[1+\left(\Gamma-1\right)\eta_1 h +
        \frac{1}{6}(\Gamma -1 ) \, \frac{a_0^2}{c_0^2}\Theta^2 \notag\\
        &
          \phantom{
               = c_*^2 \biggl[1+\left(\Gamma-1\right)\eta_1 h
          }
         - \frac{1}{3}(\Gamma - 1)(\Gamma+1) \eta_1^2 \Theta^2
   \biggr]. \label{cs}
\end{align}
Therefore, the sonic surface, $\up = \cs$, has the standard
parabolic form:
\begin{equation}
h = \frac{1}{3}(\Gamma+1)\eta_1\Theta^2.
\end{equation}

Finally, equation (\ref{cs}) enables us to estimate the spatial
derivatives of $\mu.$ Then, it can be easily shown that the terms
containing these derivatives in (\ref{GSfull'}) are negligible for
$c_0 \ll 1$ and indeed can be safely dropped. Therefore the
validity of neglecting the derivatives of $\mu$ in (\ref{GSfull'})
has been justified.

\section{Supersonic flow}
\label{Sec.SupersonicFlow} Since the pressure gradient becomes
insignificant in the supersonic region, the matter moves here
along the trajectories of free particles. Neglecting the
$\nabla_{\theta}P$ term in the $\theta$-component of relativistic
Euler equation~\citep{fro98}, we have~\citep[compare to][]{abr97}
\begin{gather}
 \alpha u_{\hat r} \frac{\partial (r u_{\hat \Theta})} {\partial r}
 + \frac{(r u_{\hat \Theta})}{r^2}
 \frac{\partial (r u_{\hat \Theta})} {\partial \Theta} +
 (u_{\hat\varphi})^2 \tan\Theta = 0.
\label{theuler}
\end{gather}
Here, using the conservation law of angular momentum,
$u_{\hat\varphi}$ can be easily expressed in terms of radius:
$u_{\hat\varphi} = 2 \sqrt{3}/r$. We also introduce dimensionless
functions $f(r)$ and $g(r)$:
\begin{align}
\Theta f(r) &= r u_{\hat \Theta}, \\
g(r) &= -\alpha u_{\hat r} > 0.
\end{align}
Using (\ref{theuler}) and the definitions above, we obtain an
ordinary differential equation for $f(r)$ which could be solved if
we knew $g(r)$:
\begin{equation}
\frac{{\rm d} f}{{\rm d} r} = \frac{f^2 + 12}{r^2 g(r)}.
\label{Eq.f}
\end{equation}

From (\ref{up2full}) we have \hbox{$\up[2] \rightarrow w^2$} as
\hbox{$r \rightarrow r_{\rm g}$}. On the other hand, $\up \approx
c_* \approx c_0$ for $r \lesssim r_*$. Therefore, the following
approximation should be valid throughout the \hbox{$r_{\rm g} < r
< r_*$} region,
\begin{equation}
g(r) \approx \sqrt{(\alpha w)^2 + (\alpha c_*)^2}.
\end{equation}

Equation (\ref{Eq.f}) governs the supersonic flow structure
for the case of a non-spinning BH. To get a better match with
observations (\cf\hbox{} Sec.\hbox{} \ref{Section.Observ}), we
also consider a more general case of a spinning BH, \ie\hbox{} a
Kerr BH with non-zero specific angular momentum $a$. After some
calculations, equation~(\ref{Eq.f}) can be generalized to the Kerr
metric with a strikingly simple form,
\begin{equation}
 \frac{{\rm d}f}{{\rm d}r} = \frac{f^2 + a^2
 \left(1-e_0^2\right)+l_0^2}{r^2 \tilde{g}\left(r\right)}.
 \label{Eq.fKerr}
\end{equation}
Here $e_0$ and $l_0$ are the specific energy and the angular
momentum of a free particle rotating at the \MSO\
respectively~\citep[see \eg\hbox{}][]{sha83}, and
$\tilde{g}\left(r\right)$ is a straightforward generalization of
$g(r)$ to the Kerr case.
For the Schwarzschild BH ($a=0$, $e_0=\sqrt{8/9}$, $l_0 =
2\sqrt{3}$) equation~(\ref{Eq.fKerr}) reduces back
to~(\ref{Eq.f}).

Integrating (\ref{Eq.fKerr}), we obtain
\begin{equation}
f\left(r\right) = \kappa \tan \left[
               \kappa \int_{r_*}^r \frac{d\xi}{\xi^{2}g(\xi)} +
                \frac{\pi}{2}
     \right],
\end{equation}
where $\kappa = \sqrt{a^2\left(1-e_0^2\right)+l_0^2}$. In the
equation above, $\pi/2$ has been to a good accuracy substituted
for the integration constant $\arctan
\left[f(r_*)/\sqrt{\kappa}\right]$: for $r$ just below $r_*,$ the
function $f$ is positive reflecting the fact that the flow
diverges; then, $f=0$ corresponds to the point where the
divergency finishes, and the flow starts to converge.

\begin{figure}
    \begin{center}
    \epsfig{figure=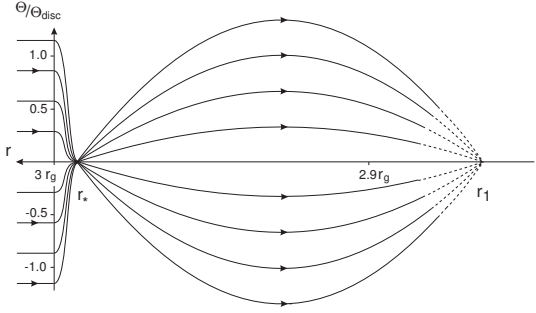,width=0.8\linewidth}
    \caption{The structure of a thin accretion
         disc (actual scale) for $c_0 = 10^{-2}$, $u_0 = 10^{-5}$
         after passing the \MSO\ $r = 3r_{\rm g}$ (Schwarzschild case).
         As sufficient dissipation
         can take place in the vicinity of the first node $r = r_{\rm 1}$,
         we do not prolong the flow lines to the region $r < r_{\rm 1}$.}
     \label{Fig.global}
     \end{center}
\end{figure}
The results of numerical calculations are presented in
Fig.\hbox{}~\ref{Fig.global}. In the supersonic region the flow
performs transversal oscillations about the equatorial plane,
their frequency independent of their amplitude. We see as well
that the maximum thickness of the disc in the supersonic (and,
hence, ballistic) region, which is controlled by the transverse
component of the gravitational force, actually coincides with the
disc thickness within the stable orbits region, $r > \rms$, where
standard estimate (\ref{2}) is correct. 

Once diverged, the flow converges once again at a `nodal' point
closer to the BH. The radial positions of the nodes are given by
the implicit formula $f(r_n)=\pm\infty,$ \ie\
\begin{equation}
\kappa \int_{r_n}^{r_*} \frac{d\xi}{\xi^{2}g(\xi)} = n \pi,
\end{equation}
where $n = 0, 1, 2, \dots$ is the node number; the node with
\hbox{$n=0$} corresponds to the sonic surface. In this formula the
sonic radius $r_* \equiv r_0$ can be to a good accuracy
approximated by $\rms = \rms(a)$ for $u_0 \ll c_0 \ll
1$~(see (\ref{r*})). The expression for $\rms(a)$ can be found in
most textbooks~\citep{sha83}.

\begin{figure}
  \begin{center}
    \epsfig{figure=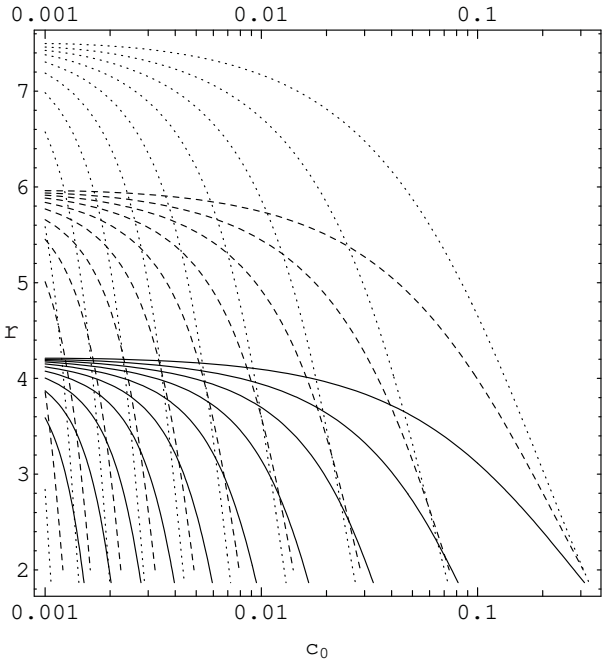,width=8cm}
    \caption{Radial positions (in the units of $M$) of the nodes
    for a range of initial sound velocities.
    Dotted, dashed, and solid curves correspond to the cases
    $a=-0.5$, $a=0$, and $a=0.5$ respectively.
    Each curve relates the radial position of a
    node to a value of the initial sound velocity. Intersection
    points of these curves with the line $c_0 = {\rm const}$ give the
    the nodes' radial positions for that particular value of $c_0$.}
     \label{Fig.nodespos}
   \end{center}
\end{figure}%

Note that equations (\ref{Eq.f}) and (\ref{Eq.fKerr}) are
inapplicable in the very vicinity of the nodes where the
pressure gradient cannot be neglected. There may also be an
additional energy release 
because the
shock fronts (and, hence, extra sonic surfaces) are inevitably to
appear there. These factors can reduce the amplitude
of the disc thickness oscillations. Accurate analysis of these factors
lies beyond the scope of this paper, and
we do not prolong the streamlines to the region $r <
r_{\rm 1}$ in Fig.\hbox{}~\ref{Fig.global}. Nevertheless, it is
quite natural to suppose that the nodes positions are not
significantly affected by the dissipative processes, and our simple approach
may give us a good qualitative understanding of the supersonic
flow structure as long as we are interested in general properties
of the flow.

Figure~\ref{Fig.nodespos} shows the positions of nodes for
different values of $c_0$ (the positions do not depend on $u_0$
for $u_0 \ll c_0$) and the BH spin parameter $a$.

Travel time between the nodes, which is in some sense similar to the 
half-period of vertical epicyclic oscillations, proves to have weak dependence 
on both $u_0$ and $c_0$. This fact provides a means for testing the theory 
via observations, and we do this in the following section.

\section{Applications to observations}
\label{Section.Observ}
\subsection{General ideas}
Our initial interest in researching into  our model's application
to observations was kindled by the publication by \citet{gen03} of
the infrared observations of the Galactic Centre (GC): one of the
flares obviously showed a number of peaks at time intervals
chirping with time (see their Fig.\hbox{}~2e), and this is exactly
what we would expect within our model.
In this section we measure physical radial distances in the units
of $GM/c^2$ and time intervals in the units of $GM/c^3$.

Suppose some perturbation in the disc (a `chunk') approaches the
\MSO. We expect to observe radiation coming from the chunk with
the period of its orbital motion,
\begin{equation}
 T_{\rm ms}\left(a\right) = 2\pi
            \left(\rms[3/2]+a\right),
 \label{Eq.Tcirc}
\end{equation}
where $a$ is the angular momentum per unit mass of the BH and
\rms\ is the estimate of the distance from the BH to the chunk.
After a number of rotations, the chunk reaches the \MSO\ and
passes through the nodal structure derived earlier (\cf\hbox{}
Sec.\hbox{}~\ref{Sec.SupersonicFlow}) generating a flare. Each
time the chunk passes through a node, it probably generates some
additional radiation, and therefore the flare is likely to consist
of several peaks. We believe that it is these peaks that were
discovered in the infrared and X-ray observations of the
GC~\citep{gen03,asc04}.

We stress that this section is quite independent of other parts of
this work. Qualitatively, our model's only assumptions are the
small thickness of the disc and the divergency of the flow in the
supersonic region (the increase of disc thickness with decreasing
radius). The latter is guaranteed provided that the flow has the
nozzle-like structure (see Sec.~\hbox{}\ref{Sec.TransFlow}) near
the sonic point.

\subsection{Finding the periods}
The time interval between the detection of two subsequent peaks
(\hbox{$n$-th} and \hbox{$(n-1)$-th}) equals the time it takes for
the chunk to pass between the two corresponding adjacent nodes
(\hbox{$n$-th} and \hbox{$(n-1)$-th}), $T_n^{(1)}$, plus
the difference in travel times to the observer for the radiation
coming from 
these two nodes,~$T_n^{(2)}$:
\begin{equation}
T_n = T_n^{(1)} + T_n^{(2)}; \label{Eq.T}
\end{equation}
$n$ can also be thought of as the index of observed time intervals
(counting from one).

The last term in (\ref{Eq.T}) is important when one of the two
nodes comes close to the BH horizon: in such a case time dilation
makes it significantly longer for the radiation to reach the
observer from that node which significantly increases $T_n^{(2)}$.
However, for outer nodes this can never happen. With this in hand,
we can estimate the value of $T_1$ neglecting the effect of
$T_1^{(2)}$. In the supersonic region, the particles move along
almost ballistic trajectories, and the orbit of a particle falling
off the \MSO\ is the circle inclined with respect to the
equatorial plane (so that the centre of the orbit is the
singularity of the BH) and perturbed by the slow radial inflow
motion. This is especially true of outermost nodes where the
matter has moderate radial velocities and time dilation is
negligible.

Suppose that at some moment in time the particle is located at the
sonic point (on the equatorial plane) which is very close from the
\MSO\ $r=\rms$ (see (\ref{r*})). Having
travelled for half the orbital period, the particle would have
reached the equatorial plane again, \ie\ the first node, therefore
\begin{equation}
 T_1 \lessapprox 1/2 \; T_{\rm ms}(a).
 \label{Eq.TnQualitative}
\end{equation}
This shows that the first time interval is a little shorter than
half the orbital period at the \MSO. Therefore, we would expect
the observations to exhibit not only the rotational frequency at
the \MSO\ but also its double counterpart, \ie\ the frequency
close to just twice of the rotational frequency.

\begin{figure}
  \begin{center}
    \includegraphics[width=\linewidth, keepaspectratio]{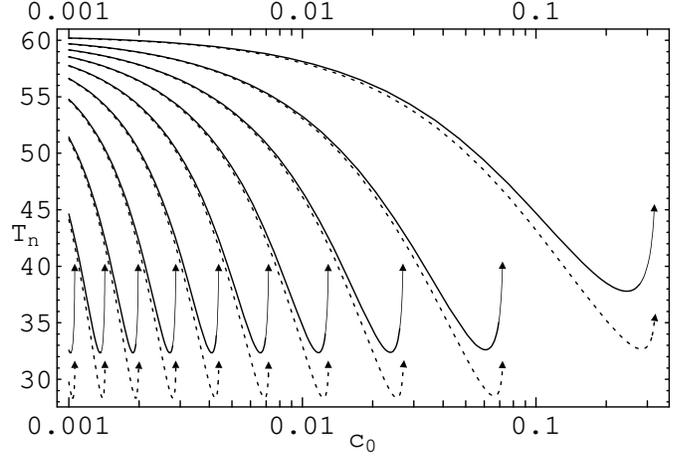}
     \caption{
      The dependence of time intervals between the peaks in a flare
      on the speed of sound in the disc, $c_0$, for a moderately spinning BH
      ($a=-0.5$); we measure time intervals in the units of $G M/c^3$.
      Solid curves correspond to $T_n = T_n^{(1)}+T_n^{(2)}$ whereas
      dashed ones correspond to $T_n^{(1)}$. The uppermost solid
      curve corresponds to the time interval $T_1$
      between the $0$th and $1$st peaks, the second solid
      curve from the top corresponds to the time interval $T_2$
      between the $1$st and $2$nd ones, etc.
      Even though individual time intervals between subsequent peaks in
      the flare may depend on the temperature in the disc (which is proportional
      to $c_0^2$, see (\ref{Tc})), their minimum and maximum values remain the
      same for the range of sound velocities where there are several intervals observed.
      In the particular case of $a=-0.5$, illustrated in the figure, we have
      $T_{\rm min} \approx 32$ and $T_{\rm max} \approx 60$
      with all other time intervals lying in between.
      For $M = 3.7 \times 10^6 M_{\bigodot}$ we obtain \hbox{$T_{\rm min} \approx 600$ s} and
      \hbox{$T_{\rm max} \approx 1100$ s}.
      }
      \label{Fig.periods}
   \end{center}
\end{figure}

Accurate calculations of $T_n^{(1)}$ -- the observer's time it
takes for the matter to travel from the $(n-1)$-th node to the
$n$-th one~-- yield%
\begin{gather}
 T_n^{(1)} \left(a, \cs\right)=
  \int_{r_n}^{r_{n-1}}
    \frac{u^t}{u^r}
    {\rm d}r
 =\int_{r_n}^{r_{n-1}}
    \frac{\left(-e_0 g^{tt} +l_0 g^{t\varphi}\right)|_{\theta=\pi/2}}
    {g\left(r\right)} {\rm d}r,
\end{gather}
where the coefficients of the inverse metric are $g^{tt} = -
\Sigma^2/(\rho^2 \Delta)$ and $g^{t\varphi} = \omega g^{tt}$; the
definitions of $\Sigma$, $\rho$, $\Delta$, and $\omega$ can be
found in~\citet{tho86}. The resulting relationship
$T_n^{(1)}(a,c_0)$ for $a = -0.5$ is shown in
Fig.\hbox{}~\ref{Fig.periods} with dashed lines.

For definiteness and simplicity, we assume that the observer is
located along the rotation axis of the BH. On its way to the
observer, the radiation travels along the null geodesic that
originates at a node in the equatorial plane (\eg\ $r=r_n$,
$\theta = \pi/2$
) and reaches the observer at infinity ($r = \infty$, $\theta =
0$). Using these as boundary conditions for
null geodesics in the Kerr metric~\citep{car68}, we numerically
find $T_n^{(2)}(a,c_0)$.

Figure \ref{Fig.periods} shows the dependence of observed time
intervals on the value of the speed of sound in the disc. All
intervals $T_n$ behave very similarly: they first decrease with
increasing $c_0$ and then abruptly increase to infinity due to
time dilation. This increase occurs if the innermost node in the
pair of nodes comes close to the BH horizon (this is indicated
with upward arrows). Although each individual time interval may
depend on $c_0$, the range [$T_{\rm min}(a)$,~$T_{\rm max}(a)$] of
observed time intervals (see the caption to
Fig.\hbox{}~\ref{Fig.periods}) is independent of $c_0$ for those
$c_0$ where there are several peaks observed. With such weak
dependence on the speed of sound in the disc, we have only two
matching parameters: the specific spin $a$ and the mass $M$ of the
BH.

\subsection{Matching the observations}
In the flare precursor section we associate the period $T_{\rm
ms}$ with the $2200 \pm 300$ s one (group 5, \cf\ Table 2 in
\citet{asc04}) and the time interval $T_1$ with the period of
$1100 \pm 100$~s (group 4 in~\citet{asc04}). In consistency with
the infrared observations of the flare, the periods $T_1$, $T_2$,
etc.\ chirp with the peak number~\citep{gen03}, \ie\ resemble the
QPO structure and thus form a cumulative peak of a larger width
shifted to higher frequencies on the flares' power density spectra
($700 \pm 100$~s, group 3, \cf\ Fig.\hbox{}~3a and 4a
in~\citet{asc04}). We can estimate the average frequency of this
peak as $1/\overline T = 1/2 \left( 1/T_{\rm min} + 1/T_{\rm
max}\right)$.

\begin{figure}
  \begin{center}
    \epsfig{figure=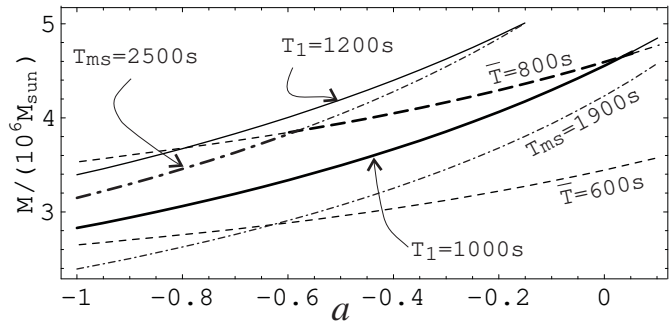,width=\linewidth}
    \caption{Relation of $M$ vs.\ $a$.
    Dashed, dash-dotted, and solid lines come from matching
    $\overline T$~(\hbox{$700 \pm 100$ s}), $T_\ms$~(\hbox{$2200 \pm
    300$ s}), and $T_0$~(\hbox{$1100 \pm 100$ s}) respectively.
    The resulting error polygon is bolded.}
    \label{Fig.aM}
  \end{center}
\end{figure}

The results of the periods' matching procedure are shown in
Fig.\hbox{}~\ref{Fig.aM}. Observational data clearly rules out
high positive values of $a$ (\ie\ the disc orbiting in the same
direction as the BH spin), and therefore the accretion disc around
the Central Black Hole is likely to be counter-rotating. Moreover,
the data presented in Fig.\hbox{}~\ref{Fig.aM} enable us to
estimate the mass of the Central Black Hole as $(3.7 \pm 1) \times
10^6 M_{\bigodot}$. This value is in agreement with the studies by
\citet{sch02} and \citet{che03}.

\section{Conclusions}
Allowing for the vertical velocity in the vertical force balance,
\ie\ performing the two-dimensional analysis, is extremely
important, especially near the sonic surface and in the supersonic
region.
Two-dimensional analysis brings about the
existence of the new small radial scale, which is of the order of
the thickness of the disc, near the sonic surface. It also shows
that the critical condition fixes neither the accretion rate nor
the angular momentum of the infalling matter.
The nozzle-like shape of the flow around the sonic point leads to
the oscillations of disc thickness in the supersonic region that
enable us to come up with a straightforward interpretation of
quasi-periodical oscillations observed in the radiation coming
from the GC \citep{gen03,asc04}. This constrains the mass of the
Central Black Hole to $(3.7 \pm 1) \times 10^6 M_{\bigodot}$ which
is in agreement with other studies \citep{sch02,che03}. Our
interpretation also suggests that the accretion disc around the
Galactic Centre black hole is counter-rotating.

\begin{acknowledgements}
We thank A.V.~Gurevich for his interest in the work and for his
support, useful discussions and encouragement. We are very
grateful to K.A.~Postnov for his help and fruitful suggestions
regarding the observational part. ADT thanks T. Elmgren for making
valuable corrections to the text.
This work was 
supported by the Russian Foundation for Basic Research (grant
no.~1603.2003.2), Dynasty fund, and ICFPM.
\end{acknowledgements}

\appendix

\section*{Appendix: Exact values of expansion coefficients}
\label{App.ExactValues}

In this section we provide exact values of expansion coefficients
(\ref{a_0})~-- (\ref{eta_3}). Here we account for the derivatives
of the metrics and keep terms of higher orders of smallness:
\begin{align}
a_0 &= \frac{\alpha_*r_*^2}{\alpha_0\rms[2]}
        \left(\frac{2}{\Gamma+1}\right)^%
        {(\Gamma+1)/2(\Gamma-1)}\frac{c_0}{u_0},
\label{a_0'}\\
a_1 &= 2 + \frac{1-\alpha_*^2}{2\alpha_*^2},
\label{a_1'} \\
a_2 &= 6a_1-6-a_1^2-(\Gamma+1)\eta_1^2,
\label{a_2'}\\
b_0 &=
 -1 - \alpha_*^2a_1^2 +
 \left(\frac{\Gamma + 1}{6}\right) \frac{a_0^2}{c_0^2}, \\
\eta_3 &=
 \frac{2}{3}\left(6a_1-6-a_1^2-\eta_1a_1
 - \frac{3}{2}\alpha_*^2a_1^2
 \right) \nonumber\\
 &\phantom{=}- \frac{\Gamma+1}{6}\frac{a_0^2-1}{c_0^2} -
 \frac{\Gamma}{3}\frac{a_0^2}{c_0^2} -
 \frac{\Gamma+1}{6}\left(1-
\frac{c_*^2r_*^2}{c_0^2\rms[2]}\right)\frac{r_*^2}{\rms[2]}
 \frac{a_0^2}{c_0^2}  \nonumber \\
 &\phantom{=}- \frac{2}{3} (\Gamma + 1)\eta_1^2,
  \label{eta_3'}
\end{align}
where $\alpha_* = \alpha(r_*)$ and $\alpha_0 = \alpha(\rms).$

\end{document}